# Latin American Studies since the 1980s: From Apocalyptic Dreams to Assimilationist Consolidation

Thiago Dumont Oliveira
*Center for Digital Text Scholarship, University of Tartu, Estonia*
Juan Grigera
*Department of International Development, King's College London, UK*
Pedro Mendes Loureiro
*Department of Politics and International Studies, University of Cambridge, UK*

## Abstract

We investigate how Latin American Studies (LAS) has evolved since 1982 in three leading LAS journals. Using network analysis, text mining, and bibliometrics, we trace transformations in disciplinary positioning, authorship, and citation patterns. Our main claim is that LAS has undergone an assimilationist consolidation. The field is now institutionally recognisable, with core LAS journals occupying central positions in disciplinary networks and acting as bridges across fields. Yet this consolidation has reinforced hierarchies privileging Anglophone academic norms. Journals based in Latin America, once central, have been marginalised. The share of Latin American authors has grown, but their profiles converge towards Northern standards.

## Introduction

The development of Latin American Studies (LAS) in Anglophone countries is traditionally seen as a by-product of the Cold War, shaped by geopolitical priorities that influenced what knowledge was produced, circulated, and legitimised, as well as who counted as legitimate interlocutors. The 1959 Cuban revolution epitomised the growing radicalisation of Latin America, with several countries electing left-wing governments in the 50s/60s, which raised a red flag in the US: 'Cuba illustrates dramatically the danger, both real and mythical, which a neighboring unfriendly government under an avowedly Communist regime represents to the people of the United States'.[1] One year prior to the revolution, Vice President Nixon and his entourage had been attacked by a 'mob' in Caracas, with Assistant Secretary of State for Inter-American Affairs Roy Rubottom stating that 'undoubtedly the attack on the Vice President was organized by the Communists'.[2] Following these events, several public and private institutions in the US 'came alive' to the 'utmost strategic importance' of Latin America to their 'national interest'.[3] The Ford Foundation distributed $270 million to area and language studies programmes between 1953 and 1966;

---

[1] Charles Wagley, 'Introduction', in Charles Wagley (ed.), *Social Science Research on Latin America* (London and New York: Columbia University Press, 1964), p. 2.

[2] U.S. Department of State, Memorandum of a Telephone Conversation Among the Minister-Counselor of the Embassy in Venezuela (Burrows), the Assistant Secretary of State for Inter-American Affairs (Rubottom) in Caracas, and the Deputy Director of the Office of South American Affairs (Sanders) in Washington, May 13, 1958, Available at: https://history.state.gov/historicaldocuments/frus1958-60v05/d46

[3] Wagley, 'Introduction', p. 13

governmental support increased considerably with the establishment of the 1958 National Defense Education Act and the launching of Project Camelot in the mid-1960s;[4] and inflation-adjusted funding to the Fulbright Program peaked in 1966.[5]

Whilst the influence of the Cold War on the development of LAS in the UK was not as overt as in the US, it certainly played a crucial role. Miller (2018) argues that the development of the field in the UK was more connected to the decline in British trade with Latin America than with geopolitical preoccupations, stressing that the Parry Report does not mention the geopolitical context.[6] However, as noted by Stansfield (1976), the Foreign Office was 'perhaps hoping to take advantage of Latin American *anti-yanquismo* generated by the Cuban revolution to recoup some of its post-war losses in the area of diplomatic influence and trade'.[7] Therefore, the Cold War created commercial opportunities which contributed to the expansion of funding for the opening of Latin American centres in the UK. Britain was the main trade partner of Latin American countries in the 19th century, but its influence declined drastically following the two World Wars and the rise of US hegemony in the region. In the early 1960s, area studies and mass media in the UK were thus mostly focused on countries that had been British colonies.[8]

Interest in Latin America also increased during the 1960s because Britain was trying to preserve its position as an important player in international affairs, following the decolonisation of the British Empire. Between 1957 and 1970, more than twenty of its former colonies gained independence. Moreover, the 1956-7 Suez crisis and Charles de Gaulle's veto of Britain's application to join the EEC in 1961 evidenced its waning imperial power. The Foreign Office reacted to the shifting geopolitical context of the 1960s by increasing its cooperation with the US, especially after Kennedy launched the Alliance for Progress in 1961: 'the expansion of Latin American Studies in the UK was undertaken at the behest of the FO [...] It was the FO that convened and co-ordinated the disparate parties sharing an interest in Latin America'.[9] The combination of commercial interests and the reconfiguration of the international balance of power thus prompted the Parry Report.

According to the Report, in 1960 there were 250 teachers working full-time on African and Asian studies, but only 9 focusing on Latin America.[10] Until the 1960s, the study of the region was restricted to sporadic institutional support from the Royal Institute for International Affairs, which published *The Republics of South America* in 1937 and a series of monographs since 1952.[11] Funding obtained via the Report would decisively change this state of affairs. The number of scholars specialising in Latin America increased from 67 to 569 between 1968 and 2009.[12] Disciplinary expertise changed substantially over this period. In 1968-9, the most represented disciplines were literature and culture (18 members) and history (19 members). By 1997, literature and culture accounted for 110 academics and history had increased to 58. In the thirty years following the report, the number of staff increased from 4 to 48 in politics, from 12 to 49 in geography, and from 6 to 74 in sociology and anthropology.[13] Between 1997-2009, literature and culture remained the most common area, increasing its presence to 153 academics, and politics assumed the second position, growing to 93. The number of historians increased to 86 – a significant growth though not as impressive as that of politics – while sociology and anthropology observed a much more modest growth, to 88. Among these subjects, only geography lost importance in absolute terms, decreasing to 43 scholars.[14]

---

4 Gabriel Paquette, 'The 'Parry Report' (1965) and the Establishment of Latin American Studies in the United Kingdom', *The Historical Journal*, 62: 1 (2019), pp. 219-240.

5 Lonnie R. Johnson, 'Memory, Commemoration, Crisis: Fulbright, Arkansas, and the Seventy-Fifth Anniversary of the Fulbright Program, 1946–2021, Part II', *Hungarian Review*, 13: 1 (2022), p. 82.

6 Rory M. Miller, 'Academic Entrepreneurs, Public Policy, and the Growth of Latin American Studies in Britain during the Cold War', *Latin American Perspectives*, 45: 4 (2018), pp. 46-68. The 'Parry Report' was a subcommittee of the University Grants Committee, a government agency responsible for distributing funds for higher education. It was established in 1962 to evaluate the state of scholarship about Latin America in Britain, and its recommendations were accepted by the government following its publication in 1965.

7 David E. Stansfield, 'The Study of Latin American Politics in British Universities', *Latin American Research Review*, 9: 2 (1974), p. 96.

8 Harold Blakemore, 'Latin American Studies in British Universities: Progress and Prospects', *Latin American Research Review*, 5: 3 (1970), pp. 111–134.

9 Gabriel Paquette, 'The 'Parry Report' (1965) and the Establishment of Latin American Studies in the United Kingdom', *The Historical Journal*, 62: 1 (2019), p. 231.

10 Blakemore, 'Latin American Studies in British Universities'.

11 Stansfield, 'The Study of Latin American Politics in British Universities', p. 96.

**'Areastreit': Latin American Studies and the Procrustean Bed of Globalism**

If the consolidation of LAS was intrinsically connected to the Cold War, the dissolution of the Soviet Union led to intense academic debates about whether area studies were still useful. Under the sign of neoliberalism, critics of the field argued that globalisation would bring about political reform and economic progress throughout the world, viz. democracy and development, thus rendering area studies obsolete and redundant. Globalisation promised transnational convergence, clearing the way for universal social science and world history.[15] The epistemological ramifications of globalisation within political science were the elevation of game theory and rational choice to the status of theoretical approaches *par excellence*, with the mainstream of the profession embracing them as scientific methods unbounded by space and able to explain a plethora of phenomena.[16]

Area Studies had come into prominence as it became widely recognised that local knowledge and methodological plurality were necessary given the structural differences of countries from different parts of the world. Such differences set limits to the external validity of theories and methods developed in the US. With the end of the Cold War, and the supposedly forthcoming effects of globalisation, these methodological boundaries dissolved in the eyes of mainstream political scientists.[17] In the words of Robert Bates, then president of the American Political Science Association, 'within the academy, the consensus has formed that area studies has failed to generate scientific knowledge'.[18] Area studies, he argued, was atheoretical and unscientific. In the academic division of labour, it played the peripheral role of providing data for political scientists to test their theories.

---

[12] For data on 1968-9, see Nikki Craske and David Lehmann, 'Fifty Years of Research in Latin American Studies in the UK', *European Review of Latin American and Caribbean Studies*, 72 (2002), p. 63. Data for 2009 are presented in Antoni Kapcia and Linda A Newson, *Report on the State of UK-Based Research on Latin America and the Caribbean* (London: Institute of Latin American Studies, 2014), p.15.

[13] Craske and Lehmann, 'Fifty Years of Research in Latin American Studies in the UK', p. 63.

[14] Kapcia and Newson, *Report*, p. 27.

[15] Jeremy Adelman, 'Latin American and World Histories: Old and New Approaches to the Pluribus and the Unum', *Hispanic American Historical Review*, 84:3 (2004), pp. 399-410.

[16] See the symposium 'Controversy in the Discipline: Area Studies and Comparative Politics' published in *PS: Political Science & Politics*, 30:2 (1997).

[17] Peter J. Katzenstein, 'Area and Regional Studies in the United States', *PS: Political Science and Politics*, 34:4 (2001), pp. 789-791.

[18] Robert H. Bates, 'Letter from the President: Area Studies and the Discipline', *APSA-CP: Newsletter of the APSA Organized Section on Comparative Politics*, 7:1 (1996), p. 1.

Once lauded as critical for strategic and military purposes, area studies entered a period of crisis. The 1990s debates about nomothetic versus idiographic approaches were reminiscent of the 'two cultures' controversy between Snow and Leavis in the 1960s[19] and of the *methodenstreit* between Menger and Schmoller in the late 19th century.[20] What differentiated the 'areastreit' from its previous iterations were the geopolitical underpinnings of its epistemological claims. The fall of authoritarian regimes and of the Soviet Union, coupled with the rise of democracies, Bates (1997) argued, 'underscored the broader relevance of the Americanists' research into elections, legislatures and political parties [... and] highlighted the broader significance of research conducted on the advanced industrial democracies'. The theories and methods developed in the US, he claimed, were now of great import for both 'the formerly socialist systems in the North and in the developing nations of the South'.[21] It was high time for comparative politics and cross-country econometric analysis.

Governmental and private agencies followed suit; the Social Science Research Council and the American Council of Learned Societies abolished area committees,[22] hundreds of area studies programmes were discontinued,[23] and the budget for the Fulbright Program was reduced by 20% in 1996 during the Clinton administration.[24] Kenneth Prewitt, then president of the SSRC, stated that area studies 'must be built on general theoretical frameworks that highlight global trends'.[25] His predecessor, David Featherman, had already advanced a similar view during his presidency (1989-95), arguing that 'disciplinary social sciences were more universally applicable, globally useful, and more worthy of support than area studies' due to their adoption of statistical methods and formal models.[26] The Ford Foundation, one of the key supporters of area studies for decades, also shifted its approach soon after Susan V. Berresford assumed its presidency in 1996. Motivated by the belief that 'some of the most pressing problems in the world aren't local or even national, but global', it discontinued its regional directorships for Africa and the Middle East, Asia, and Latin America and the Caribbean.[27]

Apart from the changing geopolitical context, LAS was also impacted during this period by broader trends affecting academic publishing. First, the availability of global datasets increased as the World Bank repositioned itself as a knowledge bank, creating opportunities for cross-country empirical analyses. Second, interdisciplinary studies, fieldwork, and ethnographic approaches require much more time and resources than empirical analysis, favouring the utilisation of publicly available datasets at the expense of methods traditionally employed by Latinistas. Third, the 1990s experimental turn led to the primacy of methods over questions. Since experiments are expensive, researchers often employ quasi-experiments and natural experiments, which require the identification of exogenous shocks. Given that such shocks are not pervasive, research questions become contingent on identifying opportunities for exploiting such methods. Fourth, the advent of the internet and the globalisation of knowledge contributed to the spread of statistical and experimental methods due to the increasing adoption of publication metrics as a proxy for research quality.[28] The success of *Designing Social Inquiry* (1994) illustrates the methodological revolution under way at the time. The book quickly became part of the toolkit of 'modern' social scientists, accruing 2,646 citations in the ten years following its publication.[29] Its stated main goal was to connect quantitative and qualitative research 'by applying a unified logic of inference to both', since 'all good research can be understood [...] to derive from the same underlying logic of inference'.[30] By laying the principles of research design, the authors intended to make 'qualitative research more scientific'.[31]

---

[19] Stefan Collini, 'Leavis v Snow: The Two-Cultures Bust-up 50 Years on', *The Guardian*, 16 Aug 2013.

[20] Uskali Mäki, 'Universals and the Methodenstreit: A Re-Examination of Carl Menger's Conception of Economics as an Exact Science', *Studies in History and Philosophy of Science Part A*, 28: 3 (1997), pp. 475-495.

[21] Robert H Bates, 'Area Studies and the Discipline: A Useful Controversy?', *PS: Political Science & Politics*, 30:2 (1997), p. 167

[22] Katja Mielke and Anna-Katharina Hornidge, 'Introduction: Knowledge Production, Area Studies and the Mobility Turn', in Katja Mielke and Anna-Katharina Hornidge (eds.), *Area studies at the Crossroads: Knowledge Production After the Mobility Turn* (New York: Palgrave-Macmillan, 2017), pp. 3-26

[23] Ian S. Lustick, 'The Disciplines of Political Science: Studying the Culture of Rational Choice as a Case in Point', *PS: Political Science & Politics*, 30: 2 (1997), pp. 175–179.

[24] Johnson, 'Memory, Commemoration, Crisis', p. 83.

[25] Peter A. Hall and Sidney Tarrow, 'Globalization and Area Studies: When is too Broad too Narrow', *Chronicle of Higher Education*, 44: 20 (1998), B4-B5.

[26] David Ludden, 'Area Studies in the Age of Globalization', *Frontiers: The Interdisciplinary Journal of Study Abroad*, 6:1 (2000), p. 5.

[27] Joye Mercer, 'The Ford Foundation Shifts its Focus and Structure', *The Chronicle of Higher Education*, August 15, 1997, A29-30.

The empirical turn in the 90s was not confined to political science. Economics, which had adopted rational choice and equilibrium analysis since the 1960s, also observed a shift from mathematical models to statistical methods in the same period.[32] This can be partly explained by the fact that empirical research is more cited than theoretical work,[33] and that citation metrics have become decisive in tenure, promotion, and distribution of resources among departments. Relying on citation metrics is problematic because they shape the perception of scholars about the quality of papers and act as a 'self-fulfilling prophecy leading to a scientific elite, which is able to reproduce its position via the mechanics of citation ranking'. They are a self-reinforcing mechanism because papers published in high-impact journals are perceived as being of high quality a priori, and, consequently, such papers are more likely to be cited, which further increases the perception that they are high-quality papers. Citation metrics thus operate as a 'conservation mechanism within science, but also have a societal role by indirectly influencing the public discourse and thus making them a hegemonial device'.[34]

Moreover, citation practices vary widely according to the methodology employed. Empirical research typically cites a high number of recent papers, boosting the impact factor of journals that focus on empirical research, whereas historians typically use archival research, older books, and government documents as sources, which do not contribute to the impact factor of history journals. The growing importance of citation metrics in the 90s was detrimental to LAS since research using ethnographic methods, interviews, archives, and historical material inevitably includes fewer references to recent articles compared to empirical research. Therefore, citation metrics implicitly define what types of research matter, making them poor proxies for research quality. Scholars interested in area studies seemed caught between Scylla and Charybdis. While publication and departmental pressure increasingly pushed them towards specialisation, the very nature of the field presupposes the ability to cross disciplinary borders. Untenured scholars were thus confronted with the choice between focusing on career development by adhering to rational choice, game theory, new institutionalism, or sophisticated statistical methods – which are of limited use to innovative scholarship in area studies – and engaging in contextualised, historical, ethnographic, critical, and interdisciplinary research – the hallmark of LAS – at the risk of not securing a permanent position in academia.

---

[28] Juan Pablo Luna, María Victoria Murillo and Andrew Schrank, 'Latin American Political Economy: Making Sense of a New Reality', *Latin American Politics and Society*, 56:1 (2014), pp. 3-10.

[29] This figure was obtained using Google scholar and filtering the period 1995-2005. The total citations to the book amounted to 16,801 as of August 2026.

[30] Gary King, Robert O. Keohane and Sidney Verba, *Designing Social Inquiry: Scientific Inference in Qualitative Research* (Princeton: Princeton University Press, 2021), pp. 3-4

[31] Ibid, p. 18

[32] Thiago D. Oliveira and Marwil J. Dávila-Fernández, 'From Modelmania to Datanomics? The Rise of Mathematical and Quantitative Methods in Three Top Economics Journals', *Scientometrics*, 123:1 (2020), pp. 51-70.

[33] David W. Johnston, Marco Piatti and Benno Torgler, 'Citation Success Over Time: Theory or Empirics?', *Scientometrics*, 95:3 (2013), pp. 1023-29.

[34] Jakob Kapeller, 'Some Critical Notes on Citation Metrics and Heterodox Economics', *Review of Radical Political Economics*, 42:3 (2010), pp. 331-2.

Although the prospects for LAS in the early years of the post-Cold War period looked grim, the neoliberal algebra of economic freedom and political stability soon proved unfounded in the face of the 1997 Asian financial crisis and the dot-com bubble in 2000. Moreover, the appeal of modernisation theory, reiterated by globalists in the 90s, dwindled with the 'war on terror' and the rise of China since 2001: 'Events serve to trip up every effort to sound the death knell of area studies', Sidaway et al. (2006) note, adding that a series of geopolitical events since the dissolution of the Soviet Union have led 'to a scramble for regional knowledge and expertise, where appropriate a lamenting of its loss, and where necessary a fillip in academic and policy interest, and in funding'.[35] In Latin America, neoliberal reforms produced an initial wave of optimism in the early 90s through renewed economic growth, lower inflation, and an acceleration of foreign investment. By 1997, however, growth and investment had slowed and dissatisfaction with the Washington Consensus was manifested in the election of several left-wing governments, starting with Hugo Chávez in 1998.[36]

In hindsight, it seems ironic that precisely when the tides were shifting in Latin America, Bates was arguing that area studies were incapable of explaining political phenomena, while the theory of games of imperfect information could fill the theoretical void since it was well-suited to explain cultural differences.[37] Regardless of the highly contested epistemic value of game theory in explaining social reality, claims about the epistemological superiority of knowledge produced in mainstream US institutions, coupled with the funding agencies' retreat from area studies, had serious implications for LAS. Although the twenty-first century would prove time and again that uneven and combined development had not been superseded by globalisation, thus vindicating the relevance of LAS, academic battles are not immediately settled by contextual factors. The attack on LAS during the 'areastreit' would persist, even if the limits of globalisation became increasingly clear, with defenders of formal methods using ad hoc explanations to motivate their models or dismissing historical events as temporary market frictions that would soften as the world converged to its new equilibrium. In the remainder of the article, we use text mining and network analysis to investigate how LAS reinvented itself to resist the pressures of limited funding and the changing academic landscape. Our main argument is that as a response to the 'areastreit', LAS has consolidated its scholarly boundaries by fully embracing interdisciplinarity. The field has become clearly distinguishable from the disciplines that contribute to it and can no longer be defined as the sum total of the distinct academic traditions that form it.

---

35 James D. Sidaway, Elaine L. E. Ho, Jonathan D. Rigg and Chih Yuan Woon, 'Area Studies and Geography: Trajectories and Manifesto', *Environment and Planning D: Society and Space*, 34: 5 (2016), p. 780.

36 Ana Margheritis and Anthony W. Pereira, 'The Neoliberal Turn in Latin America: The Cycle of Ideas and the Search for an Alternative', *Latin American Perspectives*, 34: 3 (2007), pp. 25-48.

37 Bates, 'Letter from the President' and 'Area Studies and the Discipline'.

## Data & Methods

Our data comprise 3,629 articles from *Journal of Latin American Studies* (*JLAS*), *Latin American Perspectives* (*LAP*), and *Latin American Research Review* (*LARR*) between 1982 and 2024, retrieved from Web of Science (WoS) – Table 1. We collected bibliographical references, abstracts, and institutional affiliations to assess how co-citation networks of journals, word incidence, and the participation of authors based in Latin America changed between 1982-2003 and 2004-2024. Abstracts in Spanish and Portuguese were translated into English using Google Translate; when abstracts were not available, we used the first page of articles.

**Table 1.** Article Count per Journal, 1982-2003 vs. 2004-2024

| Journal | Article Count (1982-2003) | Article Count (2004-2024) |
|---|---|---|
| JLAS | 357 | 439 |
| LAP | 562 | 1174 |
| LARR | 351 | 746 |
| Total | 1,270 | 2,359 |

In co-citation networks, journals are nodes and link strength is given by the number of times they are cited together. There is a link between two journals if they are both cited by at least one article. Therefore, journals that are more similar in scope tend to have stronger links since they are often co-cited, whereas journals with very different foci will rarely be cited by the same article. Journals can consequently have a high number of links (i.e. connect with many journals) but have a relatively low total link strength (i.e. overall number of co-citations), which makes them connectors bridging different disciplines and therefore more central in the network. Conversely, journals may have high values of total link strength even though they connect with fewer journals (i.e. low number of links), meaning they are relevant journals in terms of overall citations, but are less interdisciplinary and occupy less central positions in the network. To measure the centrality of journals we calculate their betweenness centrality,[38] which is given by:

$$C_{B(v)} = \sum_{s \neq v \neq t} \frac{\sigma_{st}(v)}{\sigma_{st}}$$

where $\sigma_{st}$ is the total number of shortest paths from node $s$ to node $t$ and $\sigma_{st}(v)$ is the number of those paths that pass through $v$.

To assess how language changed over time, we created a list of terms based on a two-step procedure. First, we computed the 50 most common words in each of the windows and checked which terms experienced significant change. Second, we created two lists with the 500 most common words in each window and then calculated the 100 words from these lists that most gained importance and the 100 words that most lost importance since the early 2000s; this step allowed us to identify several words that have lost or gained importance in relative terms, but are not amongst the 50 most common words identified in the first step. For the sake of space, we narrowed these lists by selecting terms that reflect broader disciplinary, thematic, and methodological shifts. Our final list draws on our previous discussion about the historical and academic changes taking place in the 1990s, as well as on the disciplinary shifts we have observed through our citation analysis. For example, given our discussion about globalisation and the 'areastreit', we have included in our final list the terms *global* and *transnational*, as well as a multiword term that captures the prevalence of quantitative methods within LAS.

[38] Since the number of articles has increased over time (Table 1), we have randomly selected a subset of the articles in the second window to keep the number of references equal across periods. Adjusting the article count is necessary to ensure comparability of results, since the higher number of articles in the second period makes it more likely that links exist between any pair of journals. We use the full sample for our analysis of disciplinary and thematic shifts (Figure 1 and Tables 2-3) and for the geographical distribution of authors and citations per region (Tables 5-7). For the network analysis, we used the adjusted sample keeping constant the number of references (Table 4 and Figures 2-4).

The corpus went through standard preprocessing steps: stopwords were filtered, text was converted to lowercase, and spaces, en dashes, and hyphens were normalised (e.g. 'panel data', 'panel-data', and 'panel–data' are considered the same word). Where appropriate, singular and plural forms were grouped (e.g. 'period' and 'periods'). We have also combined terms that are closely associated: the label *ethnography* includes 'ethnography', 'ethnographies', and 'ethnographic', and *territory* combines 'territory', 'territorial', and 'territories'. Finally, we have created four composite terms: *marxism*: 'marxism', 'marxist', 'marxian', and 'marx'; *gender*: 'gender', 'woman', 'women', 'feminist', 'queer', 'gay', 'lesbian', and 'trans'; *cinema*: 'film(s)' and 'cinema(s)'; and *quantitative methods*: 'statistical', 'statistic(s)', 'regression(s)', 'correlation(s)', 'probability(ies)', 'multivariate', 'logit(s)', 'logistic', 'probit(s)', 'econometric(s)', 'estimation(s)', 'measurement(s)', 'panel data', 'time series', 'fixed effect(s)', 'difference-in-differences', 'factor analysis/analyses', 'cluster analysis/analyses', 'latent class(es)', and 'simulation(s)'. We do not claim that these composite terms are perfect proxies for the relative importance of Marxism, film and gender studies, and quantitative methods; our motivation for using them is that they allow us to keep our list at a manageable size by aggregating words that share a thematic profile. In this sense, they are indicative of general disciplinary and methodological shifts even though they do not exhaust the list of issues pertaining to each of these intellectual traditions.

## Beyond 'Areastreit': Disciplinary and Methodological Changes in Latin American Studies

Table 2 presents the distribution of the 100 most cited journals by field and Fig. 1 presents the percentage of abstracts containing selected words. Taken together, Table 2 and Fig. 1 reveal considerable changes in LAS scholarship in the past two decades. The most significant change is a shift from historical analysis to political science. Considering the 100 most cited journals, history dropped from second to seventh most cited field in the second window, decreasing its presence from 14 to 5 outlets and its share of citations from 18.4% to 4.1% – see Table 2.[39] Meanwhile, the number of political science journals increased from 14 to 24, roughly doubling their share of citations (from 14.2% to 27.1%), making political science the most cited field. This shift is further confirmed by contrasting the incidence of vocabulary associated with historical analysis and political science in each period. As Fig. 1 shows, while the words *century* (−42%), *history* (−47%), and *period* (−51%) became less prevalent, *citizen* (+91%) and *governance* (+565%) grew in importance in recent decades.

Our results also indicate that Marxist and political economy scholarship lost importance in the second period. Although Marx remained among the five most-cited authors in both windows (Table 3), *New Left*

[39] Journals were labelled based on title, when unambiguous, or by the aims and scope sections in their websites. For journals that have been discontinued, information was obtained via published articles about such journals and in databases such as Latindex. Multidisciplinary journals were classified as 'other', alongside those representing disciplines that seldom appear in our list, e.g. *Science* and *The Engineering and Mining Journal.* The full list is available upon request.

*Review* dropped from 13th to 39th most cited journal, *Monthly Review* from 18th to 55th, and *Science & Society* ranked 61st in the first window, but did not figure among the 100 most cited journals in the second.[40] The share of citations to these journals fell from 3.8% to 1.5% (Table 2), and the incidence of several words that are typically used by Marxist and political economy scholarship declined considerably: *class* (−42%), *production* (−33%), *capital* (−41%), *economy* (−55%), *trade* (−52%), *capitalist* (−62%), and *Marxism* (−65%), see Fig. 1.

**Table 2:** Number of Journals and Share of Citations per Field, 1982-2003 vs 2003-2024

| Field | 1982-2003 | | 2004-2024 | |
|---|---|---|---|---|
| | Number of Journals | Share of Citations | Number of Journals | Share of Citations |
| Anthropology | 10 | 5.7% | 8 | 6.7% |
| Development | 13 | 16.8% | 12 | 13.9% |
| Economics | 5 | 3.2% | 4 | 2.6% |
| Gender Studies | 1 | 0.6% | 3 | 1.6% |
| Geography | 0 | 0% | 6 | 4.0% |
| History | 14 | 18.4% | 5 | 4.1% |
| Latin American Studies | 21 | 24.7% | 11 | 21.3% |
| Marxism | 3 | 3.8% | 2 | 1.5% |
| Political Science | 14 | 14.2% | 24 | 27.1% |
| Regional and Urban Studies | 0 | 0% | 4 | 2.4% |
| Sociology | 5 | 3.5% | 7 | 5.8% |
| Other | 14 | 9.1% | 14 | 9.0% |

These results are consistent with the fact that between 1968-1997 there were more historians than political scientists focusing on LAS, but in the period 1997-2009 political scientists assumed the second position in the ranking of disciplinary affiliations, as previously discussed.[41] Moreover, as argued by O'Donnell, since sociology departments are much older than political science departments in Latin America, most scholars of the first generation of Latinistas were sociologists. This led to 'a marked preference for macro-structural studies with an important historical dimension. The neo-Marxist and neo-Weberian approaches were very influential for this orientation'.[42] Therefore, it is not entirely surprising that history and political economy lost importance to political science in the second period.

The near absence of journals dedicated to literary and cultural studies in our lists, on the other hand, is quite surprising. As previously mentioned, literature and culture were the most represented affiliations among LAS scholars in 1997 and in 2009.[43] Yet, among the 100 most cited journals, the only journals focusing on these fields are *Casa de las Américas* and *Revista Iberoamericana*, in the first window, and *Public Culture*, in the second. In spite of the limited number of journals focusing on literary and cultural studies, our analysis of incidence of words in abstracts indicates that this trend may be changing in recent years. The word *cultural* is included in 11% of the abstracts in the second window versus 9.5% in the first window. *Memory* and *cinema*, likewise, became considerably more common in the second period, appearing in 3.2% and 2.1% of the abstracts, respectively.

[40] Moreover, although *Review (Fernand Braudel Center)* and *Cambridge Journal of Economics* are not strictly Marxist journals, both journals regularly publish Marxist scholarship and they have also dropped from the list of 100 most cited journals in the second window.

[41] See footnote 12.

[42] Guillermo O'Donnell, 'Latin America', *PS: Political Science and Politics*, 34: 4 (2001), p. 809.

[43] Kapcia and Newson, *Report*, p. 27.

**Figure 1:** Percentage of Abstracts Containing Selected Terms, 1982-2003 vs. 2004-2024

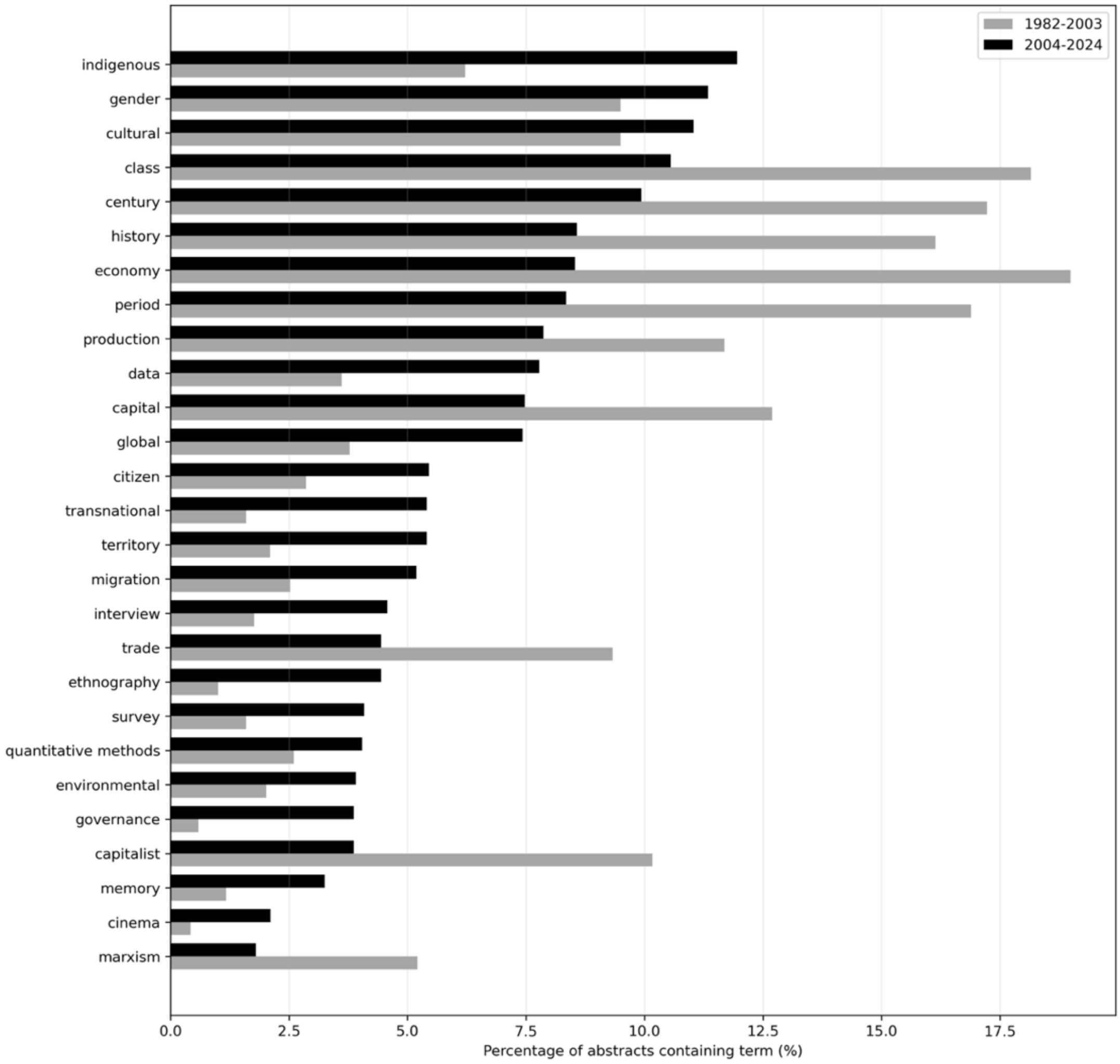

One of the factors contributing to this was Peter Ward's decision to widen the scope of *LARR* when he assumed the editorship of the journal in 2003. In his first editorial, he stressed the 'need to raise *LARR*'s visibility and role as a venue for quality research in new and emerging areas', citing cultural studies, transnationalism, and globalisation as examples of new areas.[44] Although disaggregating our results per journal is beyond the scope of our article, the effect of his editorial policies can be observed by the fact that *LARR* was the journal with the highest incidence of the word *cultural* in the second window (13.6%) versus 9% for *JLAS* and 10.2% for *LAP*. During the 80s/90s, on the other hand, abstracts containing the word were much more common in *LAP* (12.5%) than in *JLAS* (6.3%) and *LARR* (7.3%). This was likely a consequence of the heated debates between Marxists and postmodernists prompted by the 'cultural turn' in the pages of *LAP*.[45]

44 Peter M. Ward, 'Editor's Foreword', *Latin American Research Review*, 38: 1 (2003), p. 6.

45 Ronald H. Chilcote, 'Post-Marxism: The Retreat from Class in Latin America', *Latin American Perspectives*, 17: 2 (1990), pp. 3-24; Ronaldo Munck, 'Marxism, Capitalism and Socialism: Latin American Perspectives at 50', *Latin American Perspectives*, 51: 2 (2024), pp. 18-30.

Several other disciplinary, thematic, and methodological shifts occurred between 2004-2024. As shown in Fig. 1, there has been an increase in words associated with geography, anthropology, climate science, and gender studies: *ethnography* (+340%), *territory* (+158%), *environmental* (+94%), *indigenous* (+92%), and *gender* (+19%). Moreover, in the second period there were six journals of geography among the 100 most cited,[46] David Harvey became the most cited author, and Anthony Bebbington joined the list of 20 most cited authors – Table 3. In the first window, on the other hand, there were no journals of geography in the top 100 nor geographers in the top 20. Although the number of anthropology journals in our list declined from 10 to 8, their share of citations increased from 5.7% to 6.7%. Regarding gender studies, only *Signs* was amongst the top 100 journals in the first window, and it received 0.6% of the citations to these journals. In the second window, the number of journals increased to three (*Feminist Studies, Gender & Society*, and *Signs*), and their share of citations increased to 1.6% – see Table 2.

**Table 3:** 20 Most Cited Authors, 1982-2003 vs. 2004-2024

| 1982-2003 | | 2004-2024 | |
|---|---|---|---|
| Author | Citations | Author | Citations |
| O'Donnell, G. | 116 | Harvey, D. | 187 |
| Cardoso, F. H. | 108 | Bourdieu, P. | 166 |
| Marx, K. | 97 | Ellner, S. | 155 |
| Deere, C. D. | 63 | Foucault, M. | 148 |
| Knight, A. | 62 | Marx, K. | 140 |
| Portes, A. | 62 | Escobar, A. | 140 |
| Scott, J. | 61 | Levitsky, S. | 118 |
| Mesa-Lago, C. | 60 | Cardoso, F. H. | 110 |
| Petras, J. | 58 | O'Donnell, G. | 108 |
| Wolf, E. | 56 | Auyero, J. | 105 |
| Laclau, E. | 56 | Tilly, C. | 103 |
| de Janvry, A. | 55 | Robinson, W. I. | 101 |
| Brading, D. A. | 53 | Scott, J. | 100 |
| Evans, P. | 52 | Weyland, K. | 100 |
| Skidmore, T. | 50 | Svampa, M. | 100 |
| Stepan, A. | 49 | Portes, A. | 98 |
| Vilas, C. | 49 | Bebbington, A. | 95 |
| Nash, J. | 47 | Mainwaring, S. | 94 |
| Frank, A. | 46 | Gudynas, E. | 93 |
| Eckstein, S. | 46 | Quijano, A. | 92 |

These disciplinary and thematic shifts were accompanied by methodological changes. Following the 'areastreit', LAS abstracts increasingly used the words *ethnography* (+340%), *interview* (+160%), *survey*

[46] *Geoforum, Antipode, Environment and Planning A: Economy and Space, Political Geography, Annals of the Association of American Geographers, and Progress in Human Geography.*

(+156%), and *data* (+116%). However, the incorporation of *quantitative methods* was far from pervasive; in the second window, only 4% of abstracts used any of the words included in the composite term *quantitative methods*. As a reaction to mainstream political scientists' arguments about globalisation as a vindication of the universal applicability of formal and statistical methods, LAS focused its attention on the local and uneven effects of global and transnational processes. Discussions about *global* (+97%) and *transnational* (+238%) processes, such as *migration* (+106%), became much more common, but LAS diverged considerably from mainstream political science by investigating the different ways in which globalisation interacts with local realities. A cursory inspection of abstracts from our sample containing the word *transnational* illustrates how LAS incorporated these new trends.

Several articles in our sample stress the 'interplay between domestic, regional and global norms and institutions'.[47] A recurrent theme is the effect of transnational tendencies on identity construction. Scholars have explored how religious organisations have proliferated as a consequence of rapid social change, with individuals striving for 'personal security and collective identity' amidst the intense transformation of their social milieu;[48] how individuals 'negotiate the process of identity construction between the local and the global', leading to new forms of identities and citizenship as culture is redefined by these processes;[49] the ways in which indigenous peoples from different countries 'negotiate local-global networks and discourses';[50] the role of identity formation in loiceños' attempts to improve their social and economic conditions in a context of transnational struggles for resources;[51] and the differences between electoral activism in Mexico and Brazil, highlighting 'national actors' selective appropriation of transnational practices and their hybrid reinvention in national settings'.[52]

Another line of enquiry concerns the effects of the 'concentration and centralization of industrial and finance capital in powerful transnational groups';[53] including the entry of transnational agribusiness in Mexico and its deleterious effects on income inequality, land concentration, health quality, and environmental conditions;[54] the incapacity of the Mexican maquiladora industry to generate employment and dynamism in spite of its reorientation towards an export-oriented model and the increasing presence of transnational corporations;[55] the precarisation of labour conditions in Chile as a consequence of the deregulation of markets promoted by neoliberalism;[56] and the deterioration of pension systems in Latin America following the adoption of the economic reforms suggested by the World Bank in the 1990s.[57]

---

[47] Jamie Elizabeth Jacobs and Martín Maldonado, 'Civil Society in Argentina: Opportunities and Challenges for National and Transnational Organisation', *Journal of Latin American Studies*, 37: 1 (2005), pp. 141-172.

[48] Ana Celia Perera Pintado, 'Religion and Cuban Identity in a Transnational Context', *Latin American Perspectives*, 32: 1 (2005), pp. 147-173.

[49] Javier Ávila Molero and Erica Oshier, 'Worshipping the Señor de Qoyllur Ritti in New York: A Transnational Andean Ethnography', *Latin American Perspectives*, 32: 1 (2005), pp. 174-192.

[50] José Antonio Lucero, 'Representing 'real Indians': the challenges of indigenous authenticity and strategic constructivism in Ecuador and Bolivia', *Latin American Research Review*, 41: 2 (2006), pp. 31-56.

[51] Samiri Hernández Hiraldo, ''If God Were Black and from Loíza' Managing Identities in a Puerto Rican Seaside Town', *Latin American Perspectives*, 33: 1 (2006), pp. 66-82.

[52] Rafael De la Dehesa, 'Global Communities and Hybrid Cultures: Early Gay and Lesbian Electoral Activism in Brazil and Mexico', *Latin American Research Review*, 42: 1 (2007), pp. 29-51

[53] Lucio Fernando Oliver Costilla, Lilia Barragan Alvarez and Carlos Perez, 'The Reconstitution of Power and Democracy in the Age of Capital Globalization', *Latin American Perspectives*, 27: 1 (2000), pp. 82-104.

[54] Donna L Chollett, 'From Sugar to Blackberries: Restructuring Agro-Export Production In Michoacán, Mexico', *Latin American Perspectives*, 36: 3 (2009), pp. 79-92.

[55] María del Rosario Fátima Robles Robles, 'Received Wisdom: Stavenhagen's 'Erroneous Theses' and the Mexican Maquiladora Industry', *Latin American Perspectives*, 45: 2 (2018), pp. 119-128.

[56] Patricio Escobar, 'The New Labor Market: The Effects of the Neoliberal Experiment in Chile', *Latin American Perspectives*, 30:5 (2003), pp. 70-78.

[57] Camila Arza, 'Pension Reform in Latin America: Distributional Principles, Inequalities and Alternative Policy Options', *Journal of Latin American Studies*, 40:1 (2008), pp. 1-28.

A third stream of research concerns how knowledge production, cultural movements, and artistic forms in Latin America have influenced developments in the Global North. These pieces provide important examples of bilateral information flows between the South and the North which implicitly challenge arguments about the unequivocal ability of theories developed in the US to explain Latin American social, political, and economic phenomena. Examples include the influence of Cuba's revolutionary discourse on US social movements;[58] the importance of Central Americans in the development of the Central American Peace and Solidarity Movement;[59] the contribution of Brazilian sociologist and jurist Oliveira Vianna, through his engagement with the New Deal, to the insertion of 'corporatism into global debates over the role of the state in economic recovery and social welfare';[60] how 'Ni Una Menos' became a transnational movement which contributed to feminist debates in the Global North;[61] and how Donald Pierson's experience living in Brazil influenced debates at the UNESCO Race Relations Project.[62]

As these examples reveal, LAS diverged considerably from mainstream political science's predilection for empirical and formal methods in the aftermath of the 'areastreit'. As a response to globalist critiques of area studies, LAS insisted on the necessity of studying the multifarious and uneven effects of globalisation. In reaction to the empirical turn of the 90s, LAS reiterated its distinctive nature by fully embracing interdisciplinarity. This reinvention of the field can be visualised by comparing the structure of the network of journals in the two windows. Fig. 2 shows the co-citation networks of *JLAS*, *LAP* and *LARR* in 1982-2003 and in 2004-2024. For the sake of visualisation, the network contains the 100 most-cited journals. The size of nodes is proportional to their centrality, and self-citations are excluded (e.g. citations to *JLAS* in articles published in *JLAS*).

As can be seen in Figure 2 and in Table 4, LAS journals occupy central positions in the networks, especially in the second period. *Latin American Perspectives* (*LAP*) and *Hispanic American Historical Review* (*HAHR*) were the most central journals in the first window, and *Journal of Latin American Studies* (*JLAS*) and *Bulletin of Latin American Research* (*BLAR*) in the second. *Latin American Politics and Society* (*LAPS*) and *Latin American Research Review* (*LARR*) ranked third and fourth in both windows.

Betweenness centrality offers a measure of interdisciplinarity, or interconnectedness, whereas degree centrality only captures the total number of citations. For instance, although *HAHR* ranks first in citations in the first window, reaping more than three times the number of citations of *LAP*, the latter has links with a greater number of journals and scores higher in centrality than the former. Likewise, *World Development* ranks third in citations between 1982-2002, but is the fifth most central journal. This happens because it has links with 82 other journals, while *LAP*, *LARR* and *LAPS* range between 87 and 89 links.[63] Therefore, a journal may rank higher in centrality than in citations, and vice-versa. *BLAR* is the 24th most cited journal in the first window but ranks 13th in terms of centrality because it has links with a high number of journals. Conversely, *Comparative Political Studies* and *American Political Science Review* are the second and fourth most cited journals between 2002-2024, but they do not figure among the four most central journals of the network.

[58] John A. Gronbeck-Tedesco, 'The Left in Transition: The Cuban Revolution in US Third World Politics', *Journal of Latin American Studies*, 40:4 (2008), pp. 651-673.

[59] Héctor Perla Jr, 'Si Nicaragua Venció, El Salvador Vencerá: Central American Agency in the Creation of the US–Central American Peace and Solidarity Movement', *Latin American Research Review*, 43:2 (2008), pp. 136-158.

[60] Melissa Teixeira, 'Making a Brazilian New Deal: Oliveira Vianna and the Transnational Sources of Brazil's Corporatist Experiment', *Journal of Latin American Studies*, 50:3 (2018), pp. 613-641.

[61] Alyssa Bedrosian, 'How #Niunamenos Used Discourse and Digital Media to Reach the Masses in Argentina', *Latin American Research Review*, 57:1 (2022), pp. 100-116.

[62] Marcos Chor Maio and Thiago da Costa Lopes, 'Modernization, Race, and the Rural Past in Brazil: A Transnational Analysis of Donald Pierson's Sociology (1930–1950)', *Latin American Research Review*, 57:2 (2022), pp. 298-315.

[63] Centrality is not necessarily proportional to the number of links; a journal could have a small number of links but be the most central journal of a network if it is the only bridge connecting two groups of journals – see Mark S Granovetter, 'The strength of weak ties', *American Journal of Sociology*, 78: 6 (1973), pp. 1360-80. In most applications, however, nodes with more links tend to occupy more central positions in the network.

**Figure 2:** Co-citation Networks (1982-2003 vs. 2004-2024)

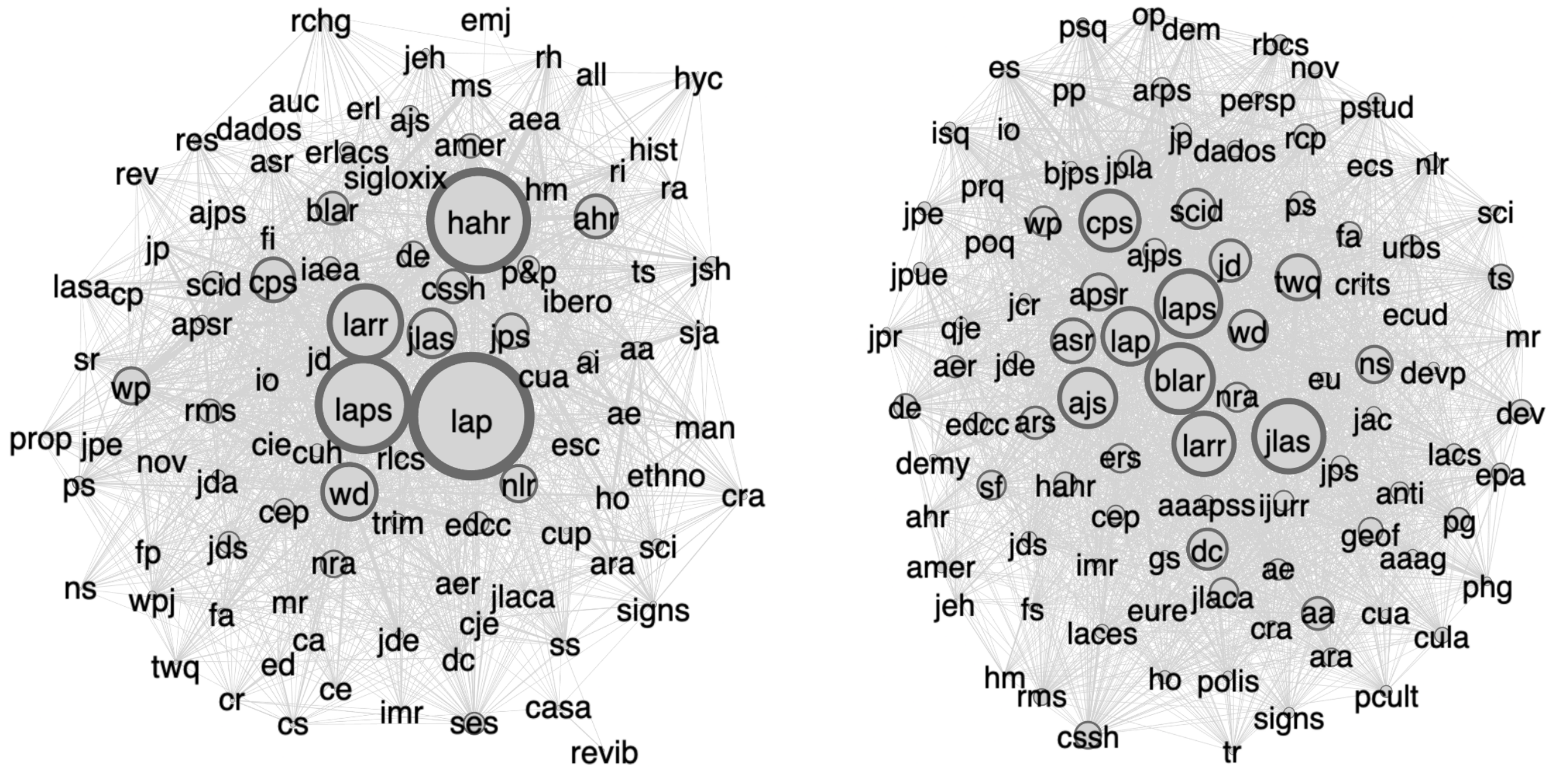

LAS journals fare better in centrality than in citations. The 10 most cited journals in the second window received 142 citations and have a centrality score of 45, on average. *LAPS*, *LARR*, *LAP*, *JLAS*, and *BLAR*, which are amongst the top 10, received virtually the same number of citations (144 on average), but their centrality score averages 58. This happens because these five journals have 95 links on average, while the other five completing the list of 10 most cited journals average 79 links. LAS journals therefore connect more widely across the network than disciplinary journals, acting as channels that facilitate information flows among neighbouring scholarly traditions.

Our results indicate that LAS has become more interdisciplinary over time not only because LAS journals rank higher in centrality in the second window relative to the first window, but also because the network has become much denser in the second period, as can be seen by visually inspecting Fig. 2. Density is a metric between 0 and 1 measuring the proportion of connections that exist relative to all possible connections that could exist. Hence, the density of a network equals 0 if no links exist between any pair of nodes and it equals 1 if all pairs of nodes are connected. Comparing the two windows, the density increased from 0.41 to 0.62. This explains why the first network has more edges and is more spatially dispersed, while the second network is smoother and more compact. These structural differences in the network indicate that research published in the three journals we investigate was more compartmentalised across disciplinary boundaries in the first window relative to the second. Due to the increase in density during the second period, the probability that two randomly selected journals are co-cited by an article picked at random is 50% higher than in the first period. Increased density therefore suggests that research has become much more interdisciplinary because the greater connectedness of the network means that journals from different disciplines are more likely to be cited together than they were in the first window.[64]

[64] The increased density of the second period also explains why centrality scores are much lower in the second period. This happens because as the number of connections increases, it becomes less likely that any randomly selected journal will be in the shortest path between any two journals. If all journals were connected, the centrality score of all journals would be 0, since none of them would be a bridge connecting other journals. Hence the more disconnected a network is, the higher centrality scores are.

**Table 4:** 10 Most Central Journals (Betweenness Centrality), 1982-2003 vs. 2004-2024

| 1982-2003 | | | 2004-2024 | | |
|---|---|---|---|---|---|
| Journal | Label | Centrality | Journal | Label | Centrality |
| Latin American Perspectives | lap | 265 | Journal of Latin American Studies | jlas | 64 |
| Hispanic American Historical Review | hahr | 218 | Bulletin of Latin American Research | blar | 61 |
| Latin American Politics and Society | laps | 202 | Latin American Politics and Society | laps | 59 |
| Latin American Research Review | larr | 160 | Latin American Research Review | larr | 56 |
| World Development | wd | 119 | Comparative Political Studies | cps | 54 |
| Journal of Latin American Studies | jlas | 103 | The American Journal of Sociology | ajs | 52 |
| Comparative Political Studies | cps | 93 | Latin American Perspectives | lap | 51 |
| American Historical Review | ahr | 91 | Third World Quarterly | twq | 40 |
| New Left Review | nrl | 77 | American Sociological Review | asr | 39 |
| World Politics | wp | 77 | Journal of Democracy | jd | 37 |

## Geographies of Knowledge: Participation of Latin American Scholars and Journals

Table 5 presents the geographical distribution of articles by journal and period. It reports the percentage of three groups of articles: those authored exclusively by researchers based in Latin America (LA), those authored exclusively by authors who are not based in Latin America (non-LA), and collaborations between LA and non-LA.[65] The percentage of articles published by LA, and of collaborations, roughly doubled in the past two decades. Taking the three journals together, the participation of LA increased from 16.2% in 1982-2003 to 32% in 2004-2024, non-LA decreased from 78.9% to 57.8%, and collaboration increased from 4.9% to 10.2%. *LARR* was the journal where the participation of LA increased the most, from 7.4% to 28.6%. This can be explained by the fact that while authors were allowed to submit their articles in English, Spanish, or Portuguese since the 1970s, such articles needed to be translated into English upon acceptance. It was only in the 2000s that *LARR* began publishing articles in the three languages, which substantially reduced costs for LA and gave them full control over the final version of the manuscript, rather than having to rely on someone else to translate it.[66]

[65] 6.9% of the articles have unknown affiliation. We exclude those articles in our calculations.

[66] Carmen Martínez Novo, 'Report from the Editor on the 60th Anniversary of LARR (2025)', *Latin American Research Review*, 60: 3 (2025), p. 490.

**Table 5:** Geographical Distribution of Articles, 1982-2003 x 2004-2024

| Journal | LA 1982-2003 | LA 2004-2024 | Non-LA 1982-2003 | Non-LA 2004-2024 | Collaboration 1982-2003 | Collaboration 2004-2024 |
|---|---|---|---|---|---|---|
| JLAS | 14.2% | 18.4% | 83.0% | 72.6% | 2.8% | 9.0% |
| LAP | 23.4% | 39.4% | 69.5% | 49.3% | 7.0% | 11.3% |
| LARR | 7.4% | 28.6% | 89.0% | 62.1% | 3.6% | 9.3% |
| Total | 16.2% | 32.0% | 78.9% | 57.8% | 4.9% | 10.2% |

Studies about the underrepresentation of Global South scholars in the production of knowledge typically investigate the geographical distribution of authors and editors of top journals.[67] While such studies are important to show that scholars from the Global North are largely responsible for producing knowledge about the Global South in leading journals, we add another dimension to this debate by arguing that it is important to consider not only where scholars are based, but also where the journals they cite are hosted. Scholars contribute directly to top journals by publishing in them and indirectly by being cited by articles published in such outlets. In this sense, the participation of knowledge produced in the Global South in top journals could increase even if the geographical distribution of authors remains stable over time if authors increasingly engage with literature produced in countries from the Global South.

Given the results presented in Table 5, one might expect that journals from Latin America have also become more prominent in the network of journals. Considering that LA engage more with Latin American journals than non-LA, as we show in what follows, the increase in participation of LA should lead to an increasing presence of Latin American journals in our network. To assess whether this has happened, Fig. 3 replicates the network presented in Fig. 2, but only Latin American journals are labelled this time.

As Fig. 3 shows, Latin American journals became less prominent in the citation networks both in terms of the number of journals that are part of the network and in terms of their positions within the network. Among the 100 most-cited journals presented in the network, the number of Latin American journals fell from 28 in 1982–2003 to 14 in 2004–2024. Moreover, their share of citations dropped from 21.2% to 10.7% over this period. Notice that Latin American journals occupied much more central positions in the network in the first window relative to the second.[68] In the first window, 6 Latin American journals were amongst the top 30 in centrality: *América Indígena* (*AI*), *Cepal Review* (*CEP*), *Desarrollo Económico* (*DE*), *Revista Mexicana de Sociología* (*RMS*), *Social and Economic Studies (SES)*, and *El Trimestre Económico* (*TRIM*). In 2004–2024, in contrast, most Latin American journals were located on the network's periphery and only 3 of them figured amongst the 30 most central journals: *Desarollo Económico* (*DE*), *Revista de Ciencia Política* (*RCP*), and *Nueva Sociedad* (*NS*).

[67] Sarah Cummings and Paul Hoebink, 'Representation of Academics From Developing Countries as Authors and Editorial Board Members in Scientific Journals: Does this Matter to the Field of Development Studies?', *The European Journal of Development Research*, 29: 2 (2017), pp. 369-383; Grieve Chelwa, 'Does Economics have an 'Africa Problem'?', *Economy and Society* 50: 1 (2021), pp. 78-99; Verónica Amarante et al., 'Underrepresentation of Developing Country Researchers in Development Research', *Applied economics letters*, 29: 17 (2022), pp. 1659-1664.

[68] A node can be in the periphery of the network and still have a relatively high centrality if it is a bridge connecting two densely connected clusters. For example, *SES* in the first window and *DE* in the second window rank higher in centrality than many Latin American journals that are positioned closer to the centre. On average, however, nodes that rank higher in centrality tend to be closer to the centre of the network; notice that nodes in the centre tend to have a bigger radius than those on the fringes of the network. Hence, the displacement of Latin American journals to the periphery of the network indicates that their centrality decreased in the second period.

**Figure 3:** Co-citation Network, LA Journals Labelled (1982-2003 vs. 2004-2024)

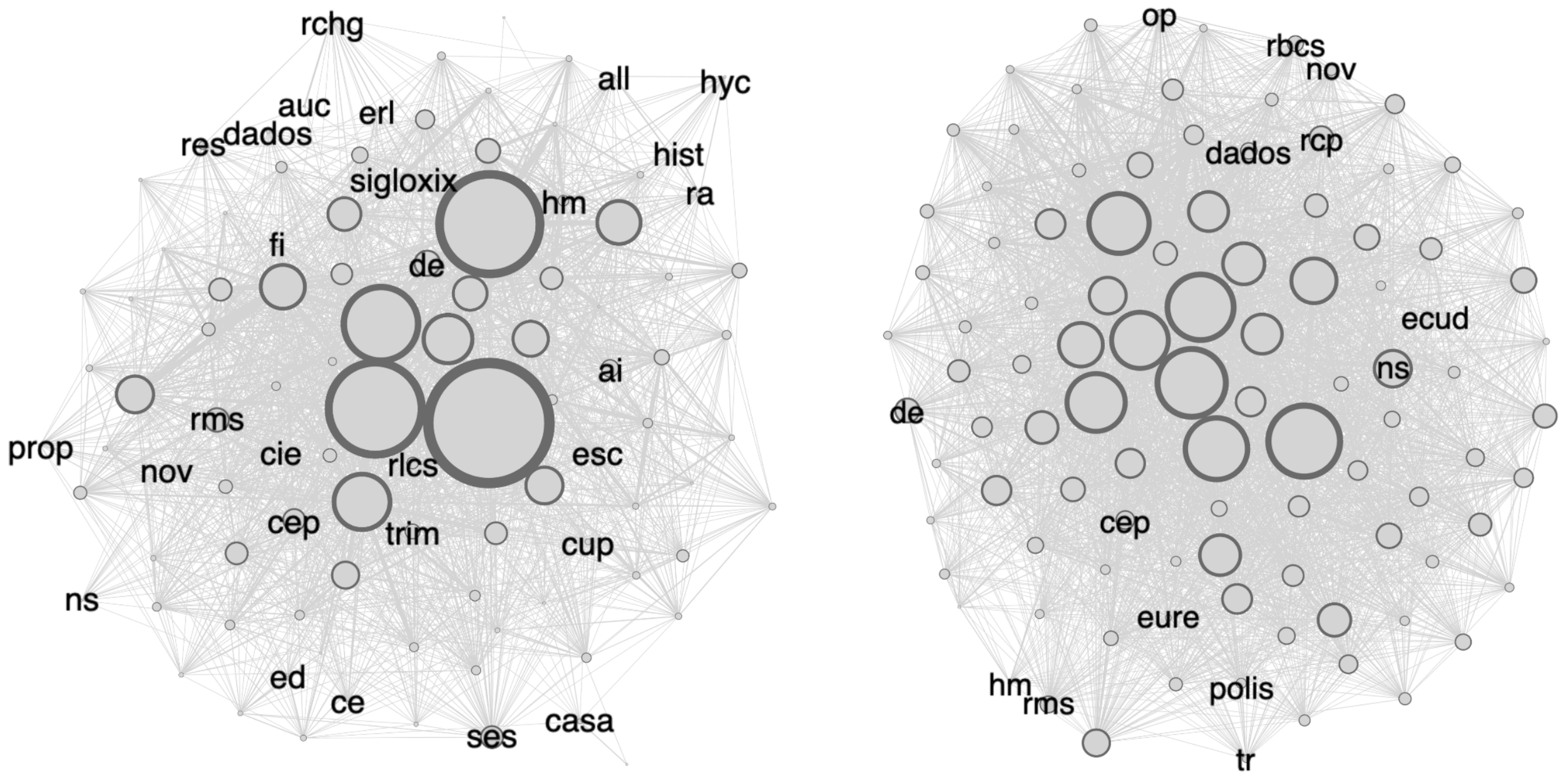

To understand why this happened, we compare the citation patterns of LA vs. non-LA. As one might expect, the former cite Latin American journals more than the latter (Tables 6-7, Latin American journals italicised). Between 1982-2003, 6 out of the 10 most cited journals by LA were from Latin America, and *Desarrollo Económico* was the most cited journal by this group. Among non-LA, on the other hand, the 10 most cited journals were hosted in UK or US institutions. This result suggests that in the 80s/90s, citation patterns varied widely between LA and non-LA. Between 2004-2024 such differences became much less visible. Although LA continued to cite Latin American journals more than scholars based elsewhere, only *Nueva Sociedad* and *Novos Estudos Cebrap* appeared amongst the 10 most cited journals. Moreover, whereas in the first window only *HAHR*, *LARR*, and *JLAS* appeared in the top 10 for both groups, in 2004-2024 the overlap between the two groups increased to 6 journals.

**Table 6:** 10 Most Cited Journals by Region, 1982-2003

| LA | Citations | Non-LA | Citations |
|---|---|---|---|
| *Desarrollo Económico* | 35 | Hispanic American Historical Review | 322 |
| Hispanic American Historical Review | 22 | World Development | 182 |
| Latin American Research Review | 20 | Latin American Research Review | 173 |
| *Novos Estudos Cebrap* | 18 | Latin American Politics and Society | 140 |
| *Revista Mexicana de Sociología* | 16 | Comparative Political Studies | 128 |
| *Nueva Sociedad* | 13 | Journal of Latin American Studies | 96 |
| *Historia Mexicana* | 12 | Latin American Perspectives | 92 |
| New Left Review | 12 | Comparative Studies in Society and History | 92 |
| *El Trimestre Económico* | 10 | NACLA Report on the Americas | 80 |
| Journal of Latin American Studies | 9 | American Political Science Review | 79 |

**Table 7:** 10 Most Cited Journals by Region, 2004-2024

| LA | Citations | Non-LA | Citations |
|---|---|---|---|
| Latin American Politics and Society | 88 | Latin American Politics and Society | 352 |
| *Nueva Sociedad* | 75 | Comparative Political Studies | 316 |
| Comparative Political Studies | 74 | Latin American Research Review | 241 |
| American Journal of Political Science | 70 | World Development | 233 |
| The Journal of Peasant Studies | 66 | American Political Science Review | 209 |
| Journal of Latin American Studies | 65 | Journal of Latin American Studies | 202 |
| Latin American Research Review | 60 | Latin American Perspectives | 193 |
| *Novos Estudos Cebrap* | 57 | Bulletin of Latin American Research | 168 |
| American Political Science Review | 57 | Hispanic American Historical Review | 165 |
| World Development | 56 | NACLA Report on the Americas | 159 |

Therefore, while LA directed most of their citations to Latin American journals in the first window, their citation practices seem to have converged towards those of non-LA over time. Latin American journals have lost prominence because they were substantially less cited by both LA and non-LA in the second period, which explains the simultaneous increase in the participation of LA and decrease in the presence of Latin American journals.

Comparing the most cited journals by the two groups is suggestive, but insufficient to assess whether citation practices converged over time. To ascertain whether this has happened, we use bibliographic coupling to calculate intergroup and intragroup link strength for the two windows. By aggregating the reference lists of all articles of each group per period, we can compute the degree of overlap between journals cited by LA and non-LA within and across windows.[69] The comparison of intergroup similarity within windows and intragroup similarity between windows allows us to determine if the groups became more similar over time, and which group changed more over time relative to its own citation practices.

Figure 4 shows that the two groups became much more homogeneous in terms of which journals they cite. The link strength between LA and non-LA groups rose from 5,571 to 10,611 in the second window, meaning that their citation profiles became substantially more similar. In principle, this could be due to LA mimicking the patterns of non-LA, or vice-versa. To understand the direction of change, it is necessary to compare intragroup similarity across time. Within-group comparisons show that there has been much more change among LA; notice that the link strength between non-LA across windows is 8,984, while it is 3,908 for LA. Therefore, articles published by non-LA have changed less over time than those published by LA. Moreover, the network shows that LA articles from 2004-2024 were more similar to non-LA articles from 1982-2003 (6,432) than to LA articles from 1982-2003 (3,908). Non-LA articles, on the other hand, were more similar to non-LA articles from the first window (8,984) than to LA articles from the first window (5,006). The comparison of between-group and within-group link strengths over time thus clearly demonstrates that LA have incorporated the citation practices of those based elsewhere, not the other way around.

[69] Link strength is given by the sum of the cross-products of citation counts for all shared journals. If an article cites 3 articles from *JLAS* and another article cites 4 articles from *JLAS*, their link strength is 12. That is why link strength is in the order of thousands although there are 187 articles and 5,147 references per group after normalisation.

**Figure 4**: Bibliographic Coupling Network, LA and Non-LA Similarity by Period

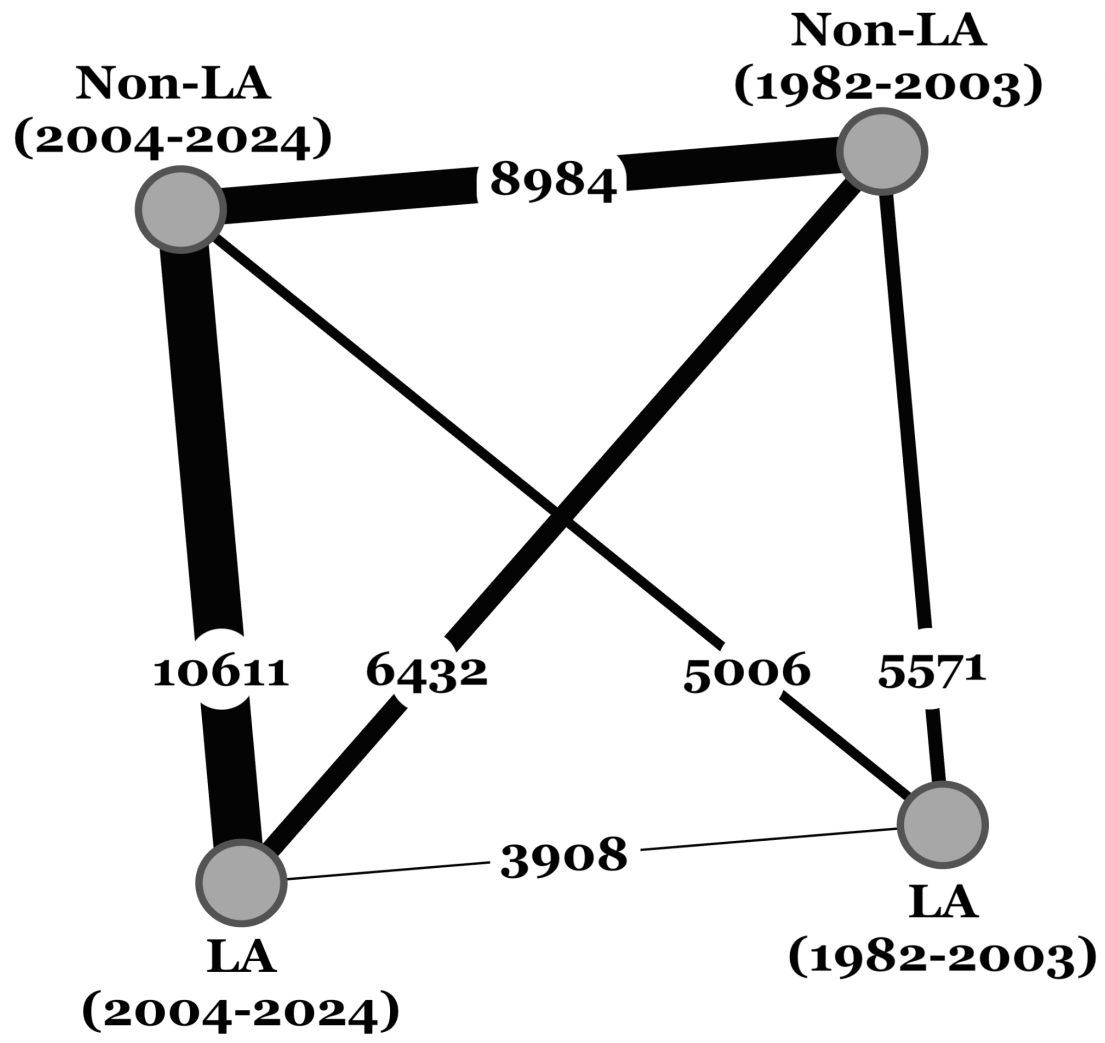

Our interpretation of these results is that by mimicking the practices of scholars based in the Global North, LA have increased their chances of publishing in leading LAS journals. Although further research would be necessary to understand this process, we argue that LAS has undergone a process of assimilationist consolidation. Scholars from Latin America increasingly participate in the production of knowledge about the region, but they draw much less on knowledge produced in Latin America. The creation of the internet has certainly contributed to the internationalisation of knowledge by increasing access to top journals for authors based in Latin America, but to reduce the trends we observe to the invention of the internet is to lose sight of the broader forces that affect both researchers in the South and in the North.

Rory Miller, reflecting on his experience as editor of *JLAS*, notes that submissions from LA have increased considerably since the early 2000s due to the advent of the internet, but that 'it was sometimes difficult to persuade a Latin American scholar writing on his/her own country that the appeal of the paper would be rather broader … if he or she framed it, at least partially, in terms of the international literature on the wide theme rather than pitching it just to a national audience'.[70] In this sense, both the research agenda of LA and the editorial policies of journals are subject to the 'citation-index mania of Northern academia', as Ronaldo Munck points out. Although no journal can 'escape that professional imperative under which all academics in most of the world now operate', he reminds us that such 'pressure can be negotiated and not just blindly reproduced'.[71]

The transformation of scholarship produced in Latin America was propelled by exogenous and endogenous forces. International funding agencies create research opportunities by funnelling resources to their preferred lines of enquiry and by refusing to fund themes that do not fall within their agenda. As forcefully put by Fernando Henrique Cardoso: 'Dependency is what we call imperialism when we apply for a Ford Foundation grant'.[72] However, Latin American countries are not merely passive recipients of international trends. They can either exacerbate or alleviate external constraints through governmental policies that set how higher education is evaluated.[73] Such policies vary by country, creating different

[70] Rory M. Miller, 'A View from the North', *Latin American Perspectives*, 43: 4 (2016), p. 117.

[71] Ronaldo Munck, 'Latin American Studies: Publishing and Politics', *Latin American Perspectives* 43: 4 (2016), p. 118.

[72] Quoted in Munck 'Latin American Studies', p. 119.

[73] Hebe Vessuri, Jean-Claude Guédon and Ana María Cetto, 'Excellence or Quality? Impact of the Current Competition Regime on Science and Scientific Publishing in Latin America and its Implications for Development', *Current Sociology* 62: 5 (2014), pp. 647-665; Fernanda Beigel, et al. 'OLIVA: The Scientific Output in Journals Edited in Latin America. Disciplinary Diversity, Institutional Collaboration, and Multilingualism in Scielo and Redalyc (1995-2018)', *Dados* 67: 1 (2024); Bernhard

career progression requirements that affect scholars' decisions about where to submit their articles, whether to publish them in English or in their native language, and what methods, theories, and approaches they should use.

In Ecuador, for example, achieving the highest salary scale as a full professor effectively requires scholars to publish in non-Ecuadorian journals and severely limits the possibility of publishing in Spanish, given the high standards the country imposes in terms of the number of articles published in journals ranked level 1 and the criteria it adopts for what counts as indexed journals. Journals from smaller countries like Ecuador, which are less likely to be indexed in international databases, have suffered the consequences of the stricter policies enacted during Rafael Correa's administration,[74] but the increasing importance of publishing in English and in top journals has affected most countries in Latin America.[75] As shown by Lucía Céspedes, articles published in Spanish and Portuguese, taken together, represented 3% of the Scopus database and 2.7% of Web of Science in 2020. This happened not only because Latin American journals are poorly represented in these databases, but also because 35% of articles published in Latin American journals indexed in Web of Science were published exclusively in English, and 13% of those indexed in Scopus.[76]

The diminishing importance of Latin American journals, reported in Figure 3, was thus a consequence of both international trends in academic publishing and national governments' policies regarding research evaluation and career progression requirements. Our results indicate that following the 1990s 'areastreit', LAS went through a process of assimilationist consolidation. It has stabilised as a field by embracing interdisciplinarity and forging an organic unity that is more than the amalgamation of coterminous disciplinary traditions, but it has done so largely through the assimilation of scholars based in Latin America into Anglophone academic hierarchies rather than through genuine pluralisation.

## Conclusion

Some of the main LAS journals have recently published pieces celebrating milestone anniversaries and discussing the main achievements of the field in the last 50-odd years. In 2024 *LAP* celebrated its 50th anniversary, and 2025 marked the 60th anniversary of *LARR* and of the *European Review of Latin American and Caribbean Studies* (*ERLACS*). We close our article by assessing three points raised in these reflections in light of our results: interdisciplinarity, recent trends, and dialogue with scholarship from Latin America. While our findings support some of the arguments recently put forth about the evolution and recent state of LAS, they also reveal some of these assessments to be overly optimistic.

One of the recurrent themes in reflections about LAS concerns its commitment to interdisciplinarity as one of its distinguishing features. In a special issue commemorating 50 years of *LAP*, Chilcote stresses that since its outset the editorial collective's 'vision for the journal was broad, reaching beyond our academic disciplines and particular field experiences'. In his estimation, it has fulfilled this vision by offering an outlet for various perspectives at 'the intersection of class, gender, race, and ethnicity' that are

---

Sabel and Dan Larhammar, 'Reformation of Science Publishing: The Stockholm Declaration', *Royal Society Open Science*, 12: 11 (2025).

74 Carmen Diana Deere, 'Publishing in International, Peer-reviewed Social Science Journals: Implications for Latin American Studies', *Latin American Perspectives*, 43: 4 (2016), pp. 107-111

75 Magdalena Suárez-Ortega, Elisa García-Mingo and José A. Ruiz San-Román, 'When Español Is Not Enough: Research, Write, Translate and Publish or… Perish', *International journal of leadership in education*, 15: 4 (2012), pp. 463-482; Fernanda Beigel, 'Publishing from the Periphery: Structural Heterogeneity and Segmented Circuits. The Evaluation of Scientific Publications for Tenure in Argentina's CONICET', *Current Sociology*, 62: 5, (2014), pp. 743-765; Elena Sheldon, ''We Cannot Abandon the Two Worlds, we Have to Be in Both' Chilean Scholars' Views on Publishing in English And Spanish', *Journal of English for Research Publication Purposes*, 1: 2 (2020), pp. 120-142; Kevin Aguirre-Carvajal and Vinicio Armijos-Jaramillo, 'Global Integration, Local Constraints: How Internationalized Scientific Development Shapes Early-Career Research in Latin America', *Frontiers in Sociology*, 11 (2026).

76 Lucía Céspedes, 'Latin American Journals and Hegemonic Languages for Academic Publishing in Scopus and Web of Science', *Trabalhos em Linguística Aplicada* 60: 1 (2021), pp. 141-154.

‘critical of narrow institutional analysis’.[77] In an earlier article, he linked the interdisciplinarity of the field with the scholarship of Latin Americans who ‘contributed new ideas and theoretical understandings, and LAS evolved in interdisciplinary ways and through collaboration to transcend barriers of disciplinary intransigence’.[78] In her report on the 60th anniversary of *LARR*, Novo also highlights the interdisciplinarity of the journal, touching on some of the pressures regarding academic publishing that we have discussed in the article. She notes that during the editorship of Aníbal Liñán (2016-2020), the journal was organised by disciplines in an attempt to improve its impact factor, given that disciplinary journals are more cited than area studies. Under Novo’s editorship, however, ‘the journal has returned to its traditional interdisciplinary emphasis. Articles are no longer organized by disciplines, but as thematic conversations’.[79] Likewise, in their editorial celebrating 60 years of *ERLACS*, Hogenboom and Fernández Rodríguez stress that ‘the contributions in this special issue share a common message: through their forward-looking perspectives, they point to what can be gained from establishing new or stronger connections among themes and/or disciplines’.[80]

Combining bibliometrics and network analysis, we have shown not only that LAS journals are more interdisciplinary than subject-specific journals, as argued in recent assessments of the field, but also that they have become more interdisciplinary following the ‘areastreit’. LAS journals occupy central positions in the networks in both periods, particularly in the second window. Between 2004-2024, the four most central journals of the network were *JLAS*, *BLAR*, *LAPS*, and *LARR*. Moreover, the density of the network increased by roughly 50% in the second period, indicating that research became less compartmentalised by disciplines. The increased connectedness of the network indicates that two randomly selected journals of the network were more likely to be co-cited in the second window; hence ideas originating from different disciplines or different traditions within the same disciplines were more likely to co-occur. As we have shown, LAS journals rank better in terms of betweenness centrality than in terms of degree centrality. By establishing links with a large number of journals, they occupy central positions in the network and act as bridges between journals that do not directly communicate with each other. Therefore, although LAS journals are less cited than some disciplinary journals in our sample, they have more edges than such journals and play the role of connectors across neighbouring scholarly communities.

A second issue that has been discussed in recent articles about LAS concerns how the field has changed in recent years. Hogenboom and Fernández Rodríguez, analysing the 60 most downloaded articles of *ERLACS* in the past 10 years, highlight that the five topics of research that most attracted the attention of the LAS community were ‘politics, democratic retreat and authoritarianism; extractivism; migration and mobility; gender and sexuality; and indigenous peoples and territorial justice’.[81] In the same special issue, Castiglioni and Ibarra González analysed 801 articles in 10 LAS journals between 2022-2024. They identify six broad areas commonly explored: ‘1) procesos sociales; 2) problemas político-institucionales; 3) géneros y feminismos; 4) políticas públicas y sociales; 5) etnia e indigenismo; y 6) trabajo y condiciones laborales’. Apart from these broader areas, the authors highlight four topics that are often addressed by LAS scholars, namely climate change, transnational organised crime, democratic backsliding, and migration.[82] Novo notes that *LARR* has striven to promote diversity by publishing ‘widely on race, ethnicity, racism, gender, and sexuality’, and that the journal has ‘prioritized current issues like democratic backsliding, human rights, the struggles for memory, and the advance of authoritarianism and the far right'.[83] Munck argues that apart from *LAP*'s traditional focus on the study of class struggle, the capitalist state, and the transition to socialism, the journal has actively contributed in recent years towards building a new research agenda for LAS that critically engages with 'gender, intersectionality, identity formation, culture (including religion), and sustainable development'.[84]

---

[77] Ronald H Chilcote, ‘Our Mission as a Journal’, *Latin American Perspectives*, 51: 2 (2024), pp. 7-8.

[78] Ronald H Chilcote, ‘The Cold War and the Transformation of Latin American Studies in the United States’, *Latin American Perspectives*, 45: 4 (2018), p. 37.

[79] Novo, ‘Report from the Editor’, p. 491.

[80] Barbara Hogenboom and Rebeca Fernández Rodríguez, ‘Editorial Afterword: Celebrating 60 Years of Publishing ERLACS Trends and Gaps in Latin American and Caribbean Studies’, *European Review of Latin American and Caribbean Studies*, 120 (2025), p.114.

[81] Hogenboom and Fernández Rodríguez, ‘Editorial Afterword’, p. 112.

[82] Rossana Castiglioni and Guzmán Ibarra González, ‘Mapeando los Estudios de Área sobre América Latina: Características Críticas y Desafíos’, *European Review of Latin American and Caribbean Studies*, 120 (2025), p. 103.

The disciplinary and thematic shifts we have identified provide empirical support for these reflections about recent trends in LAS. As our comparison of the 100 most cited journals in 1982-2003 vs. 2004-2024 has shown, political science has become a major source for LAS in the past two decades, with journals of the discipline accruing roughly a quarter of the citations to the top 100 journals. Meanwhile, history was a leading force in the first period, but lost importance in the second. Accordingly, the articles by Hogenboom and Fernández Rodríguez and by Castiglioni and Ibarra González highlight political issues as central areas of research within LAS scholarship but make no reference to historical analysis. Both articles, as well as Munck's, also mention gender studies as an increasingly important topic for LAS. This is consistent with our results; in the second window the number of journals within the top 100 dedicated to gender studies (and their share of citations) increased from 1 (0.6%) to 3 (1.6%). Moreover, using text mining we have shown that the percentage of abstracts containing *gender* increased from 9.5% to 11.3%. Overall, the results of our textual analysis overlap significantly with the recent trends outlined in the previous paragraph. As we have shown, themes whose incidence in abstracts increased substantially in the second window include *ethnography* (+340), *transnational* (+238%), *territory* (+158%), *migration* (+106%), *global* (+97%), *environmental* (+94%), and *indigenous* (+92%). Finally, our result that the proportion of abstracts including *cultural* has increased from 9.5% to 11% supports Munck's suggestion that identity formation and culture are critical themes for the future of LAS.

The final point we would like to discuss concerns the participation of scholars from Latin America in the production of knowledge. In Chilcote's reconstruction of the transformation of LAS during the Cold War, he concludes that LAS in the US made advancements largely by 'engaging with debate within Latin America', adding that the field recognised 'that intellectual endeavor and scholarly contributions necessitate Latin American perspective and participation'.[85] Castiglioni and Ibarra González, likewise, argue that since its outset LAS has promoted collaboration with scholars from Latin America: 'estos estudios han promovido la colaboración entre académicas y académicos radicados en América Latina y quienes residen en el norte global … los investigadores del norte trabajan *con* sus contrapartes latinoamericanas, y no *para* ellas … los estudios de área latinoamericanos han desafiado el centralismo académico norteamericano, generando conocimientos *desde* la región, de forma cada vez más horizontal y descentralizada'.[86]

Our analysis casts doubt on whether the production of knowledge has become increasingly decentralised and horizontal. Although the proportion of authors from Latin America has roughly doubled in the last twenty years, the number of journals from the region within the top 100 and their share of citations have roughly halved. The increasing participation of authors from Latin America was thus offset by the decrease in citations to journals from the region. Moreover, we have shown that such journals have been pushed to the periphery of the network. This happened despite the increasing proportion of authors from Latin America, because those authors' engagement with Latin American journals has decreased. The comparison of intergroup and intragroup link strength shows that Latin American scholars have mimicked the citation practices of scholars from other countries.

Taken together, our results indicate that LAS has undergone an assimilationist consolidation. The field is now institutionally recognisable, with core LAS journals occupying central positions in disciplinary networks and acting as bridges across fields. Yet this consolidation has reinforced hierarchies privileging Anglophone academic norms. Journals based in Latin America, once central, have been marginalised. The share of Latin American authors has grown, but their profiles converge towards Northern standards. These results are quite concerning, especially given the increasing use of AI in research. LLMs reproduce societal stereotypes and they may reinforce scientific inequities by referencing works of prominent male scholars from the Global North to a much higher degree than those from Latin America. Given that these epistemological biases make it more challenging to increase the participation of scholars from Latin America in the production of knowledge about their region, the design and implementation of policies that mitigate such problems are more important than ever.

[83] Novo, 'Report from the Editor', p. 491.

[84] Munck, 'Marxism, Capitalism and Socialism', p. 27.

[85] Chilcote, 'The Cold War and the Transformation of Latin American Studies in the United States', p. 37.

[86] Castiglioni and Ibarra González, 'Mapeando los Estudios de Área sobre América Latina', p. 101.